\documentclass[Physsubmission, Phys]{SciPost}

\hypersetup{
    colorlinks,
    linkcolor={red!50!black},
    citecolor={blue!50!black},
    urlcolor={blue!80!black}
}

\newcommand{\pT}{\ensuremath{p_{\rm{T}}}}
\newcommand{\pp}{$pp$}
\newcommand{\AuAu}{Au$+$Au}
\newcommand{\zg}{$z_{\rm{g}}$}
\newcommand{\rg}{$R_{\rm{g}}$}

\usepackage[bitstream-charter]{mathdesign}
\DeclareSymbolFont{usualmathcal}{OMS}{cmsy}{m}{n}
\DeclareSymbolFontAlphabet{\mathcal}{usualmathcal}

\begin{document}

\begin{center}{\Large \textbf{
Measurement of splittings along a jet shower in \\$\sqrt{s} = 200$ GeV \pp\ collisions at STAR\\
}}\end{center}

\begin{center}
Raghav Kunnawalkam Elayavalli\textsuperscript{1,2} for the STAR Collaboration
\end{center}

\begin{center}
{\bf 1} Wright Laboratory, Yale University
\\
{\bf 2} Brookhaven National Laboratory
\\
raghav.ke@yale.edu
\end{center}

\begin{center}
\today
\end{center}


\definecolor{palegray}{gray}{0.95}
\begin{center}
\colorbox{palegray}{
  \begin{tabular}{rr}
  \begin{minipage}{0.1\textwidth}
    \includegraphics[width=22mm]{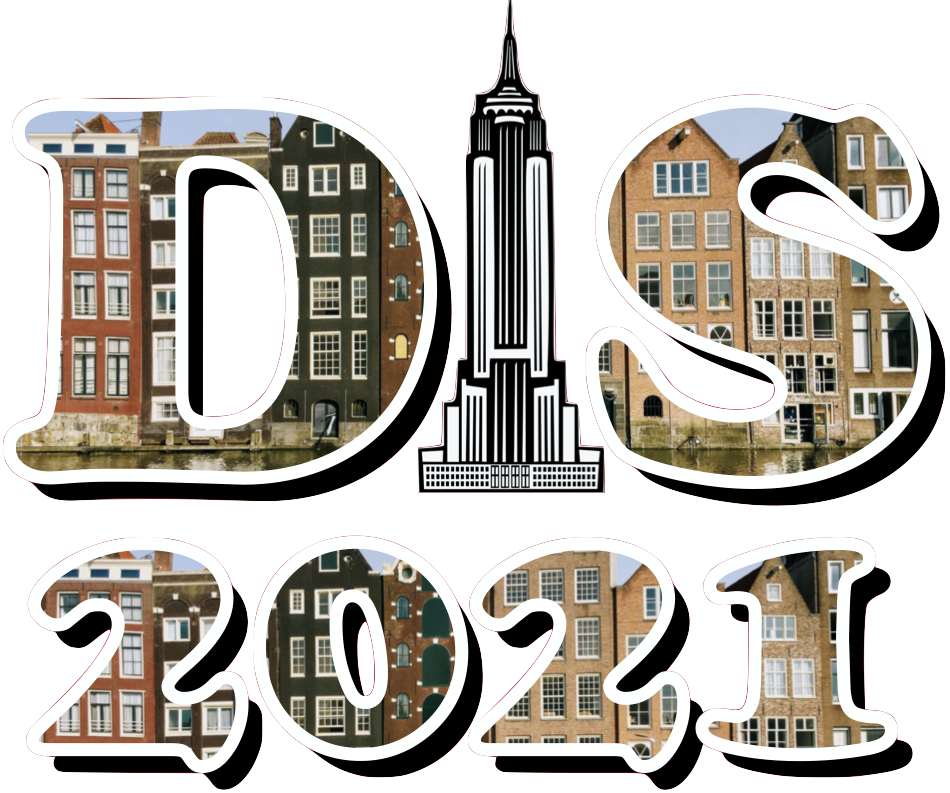}
  \end{minipage}
  &
  \begin{minipage}{0.75\textwidth}
    \begin{center}
    {\it Proceedings for the XXVIII International Workshop\\ on Deep-Inelastic Scattering and
Related Subjects,}\\
    {\it Stony Brook University, New York, USA, 12-16 April 2021} \\
    \doi{10.21468/SciPostPhysProc.?}\\
    \end{center}
  \end{minipage}
\end{tabular}
}
\end{center}

\section*{Abstract}
{\bf Jets are algorithmic proxies of hard scattered partons, i.e. quarks and gluons, in high energy particle collisions. The STAR collaboration presents the first measurements of substructure observables at the first, second and third splits in the jet clustering tree via the iterative SoftDrop procedure. For each of these splits, we measure the fully corrected groomed shared momentum fraction (\zg) and groomed jet radius (\rg). We discuss the evolution of jet substructure in both the angular and momentum scales which allows for a self-similarity test of the DGLAP splitting function. We compare the fully corrected data to Monte Carlo models, providing stringent constraints on model parameters related to the parton shower and non-perturbative effects such as hadronization.}

\section{Introduction}
\label{sec:intro}

Jets are composite objects resulting from a convolution of parton shower (perturbative-QCD) and fragmentation (non-perturbative-QCD) processes, and as such they contain rich substructure information that can be exploited via jet finding algorithms~\cite{Marzani:2019hun}. These algorithms typically employ an iterative clustering tree procedure that generates a tree-like structure, which upon an inversion, gives access to jet substructure at different steps along the cluster tree. The most common toolkit for such measurements is SoftDrop~\cite{Larkoski:2014wba} which employs a Cambridge/Aachen re-clustering of jet constituents and imposes a criterion at each step as we walk backwards in the de-clustered tree, 
\begin{equation}
\label{eq:zg}
z_{g} = \frac{\rm{min}(\it{p}_{\rm{T, 1}}, p_{\rm{T, 2}})}{p_{\rm{T, 1}} + p_{\rm{T, 2}}} > z_{\rm{cut}} \left( \frac{R_{\rm{g}}}{R_{\rm{jet}}}\right)^{\beta}; ~ ~ R_{\rm{g}} = \Delta R (1, 2),   
\end{equation}
where $1, 2$ are the two prongs at the current stage of de-clustering, $p_{\rm{T}}$ is the transverse momentum of the respective prong, $R_{\rm{jet}}$ is the jet resolution parameter and $\Delta R$ is the radial distance in the pseudorapidity $\eta$-azimuthal angle($\phi$) plane. The free parameters in Eq.~\ref{eq:zg} are $z_{\rm{cut}}$ a momentum fraction threshold, and $\beta$, the angular exponent which in our analysis are set to $0.1$ and $0$, respectively~\cite{Larkoski:2015lea}. These parameter values make SoftDrop observables calculable in a Sudakov-safe manner, and at the infinite jet momentum limit they converge to the DGLAP splitting functions.  
STAR recently measured the SoftDrop groomed shared momentum fraction (\zg) and groomed jet radius (\rg) at the first surviving split for jets of varying transverse momenta and jet radii~\cite{STAR:2020ejj}. These double differential measurements were fully corrected in both jet \pT\ and \zg/\rg\ simultaneously. The data demonstrate a significant variation in \rg\ as the $p_{\rm{T,jet}}$ increases, reflecting momentum dependent narrowing of jet substructure, whereas \zg\ only varies slowly and has a relatively constant shape for $p_{\rm{T,jet}} > 30$ GeV/$c$. 

Since the jet clustering tree extends beyond the first split, we iteratively apply the SoftDrop procedure on the hardest (highest \pT) surviving branch and measure the jet substructure at each split along the de-clustered tree~\cite{Dreyer:2018tjj}. Such measurements enable, for the first time, a time-differential study of the parton shower and evolution of both the momentum (\zg) and angular scales (\rg) within a jet. Upon applying the iterative SoftDrop procedure, with the same aforementioned values of the parameters, we reconstruct a collection of observables corresponding to $z^{n}_{\rm{g}}$ and $R^{n}_{\rm{g}}$ at a given split $n$. We limit our measurement to the first three surviving splits within each jet and present the results fully corrected in 3-D corresponding to the jet or initiator \pT, $z_{\rm{g}}/R_{\rm{g}}$, and the split number $n$ for jets of varying $p_{\rm{T, jet}}$ and for splits of varying initiator \pT. This provides the potential benefit of studying the self-similarity of the QCD splitting functions. 

\section{Analysis details}
\label{sec:star}

The \pp\ data utilized in this measurement was collected with the STAR detector~\cite{star} during the 2012 run at $\sqrt{s} = 200$ GeV.  Events are selected by an online jet patch trigger in the Barrel ElectroMagnetic Calorimeter (BEMC) which is a $1\times1$ patch in $\eta \times \phi$ with a total sum $E_{\rm{T,patch}} > 7.3$ GeV. Events are also required to have their primary vertices, reconstructed via charged particle tracks from the Time Projection Chamber (TPC), to be within $|v_{\rm{z}}| < 30$ cm along the beam axis from the center of the detector. Jets are reconstructed from charged tracks ($0.2 < p_{\rm{T}} < 30.0$ GeV/$c$) in the TPC and energy depositions in the BEMC towers ($0.2 < E_{\rm{T}} < 30.0$ GeV) using the anti-$k_{\rm{T}}$ algorithm with a resolution parameter $R_{\rm{jet}} = 0.4$ as implemented in the FastJet package~\cite{FastJet}. Same track, tower and jet selections are applied as in~\cite{STAR:2020ejj}. 

A novel correction technique is employed for this 3-D measurement. Detector smearing effects on the substructure observables \zg\ and \rg\ at a given split, and at a given initiator \pT\ or jet \pT\ are unfolded via a 2-D Iterative Bayesian procedure as implemented in the RooUnfold package~\cite{DAgostini:1994fjx}. The detector response is estimated via \textsc{PYTHIA 6} (Perugia 2012 tune~\cite{Sjostrand:2006za} and further tuned to STAR data~\cite{STAR:2018yxi}) events passed through a \textsc{GEANT3} simulation of the STAR detector. These simulated events are embedded into zero-bias \pp\ data and the resulting events are analyzed in a similar fashion to the real data. Since the splits are identified at the detector level, detector effects on the jet clustering tree could mangle the split hierarchy, i.e. splits at the particle level can be lost or mis-categorized in the detector-jet clustering tree, along with the addition of fake splits arising from particles of uncorrelated sources, such as interactions with detector material. To correct the split hierarchy, we introduce an additional matching requirement of the splits based on the initiator prong at the particle and detector-level via $\Delta R (\rm{initiator}_{\rm{det, part}}) < 0.1$ to build  a hierarchy matrix with particle-level splits on the $x-$axis and detector-level splits on the $y-$axis. The 2-D unfolded data are then added with the relevant weights along each column of the hierarchy matrix to get a fully corrected particle-level distribution of \zg\ and \rg\ as a function of the jet/initiator \pT\ at a true split $n$. 

The systematic uncertainties follow the same procedure outlined in~\cite{STAR:2020ejj}, and are broadly grouped into two categories: detector performance and analysis procedure. The former sources of uncertainties constitute variations of the tracking efficiency by $\pm 4\%$ and tower energy scale by $\pm 3.8\%$.  The systematic uncertainty due to the analysis procedure includes hadronic correction, i.e.~correcting $100\%$ to $50\%$ of the matched track's momentum from a tower's energy to negate double counting of energy depositions. Uncertainty due to the unfolding procedure is taken as the maximal envelope of variations in the iteration parameter and shape uncertainties arising from the prior (varied by the differences to \textsc{PYTHIA 8}~\cite{Sjostrand:2014zea} and \textssc{HERWIG 7}~\cite{Bellm:2015jjp}). Lastly, the split matching criterion is varied by $\pm 0.025$ and the consequent variation to the fully corrected result is taken as a shape uncertainty. 

\section{Results}
\label{sec:results}

\begin{figure}[h]
\centering
\includegraphics[trim=120 60 0 20,clip,width=0.7\textwidth]{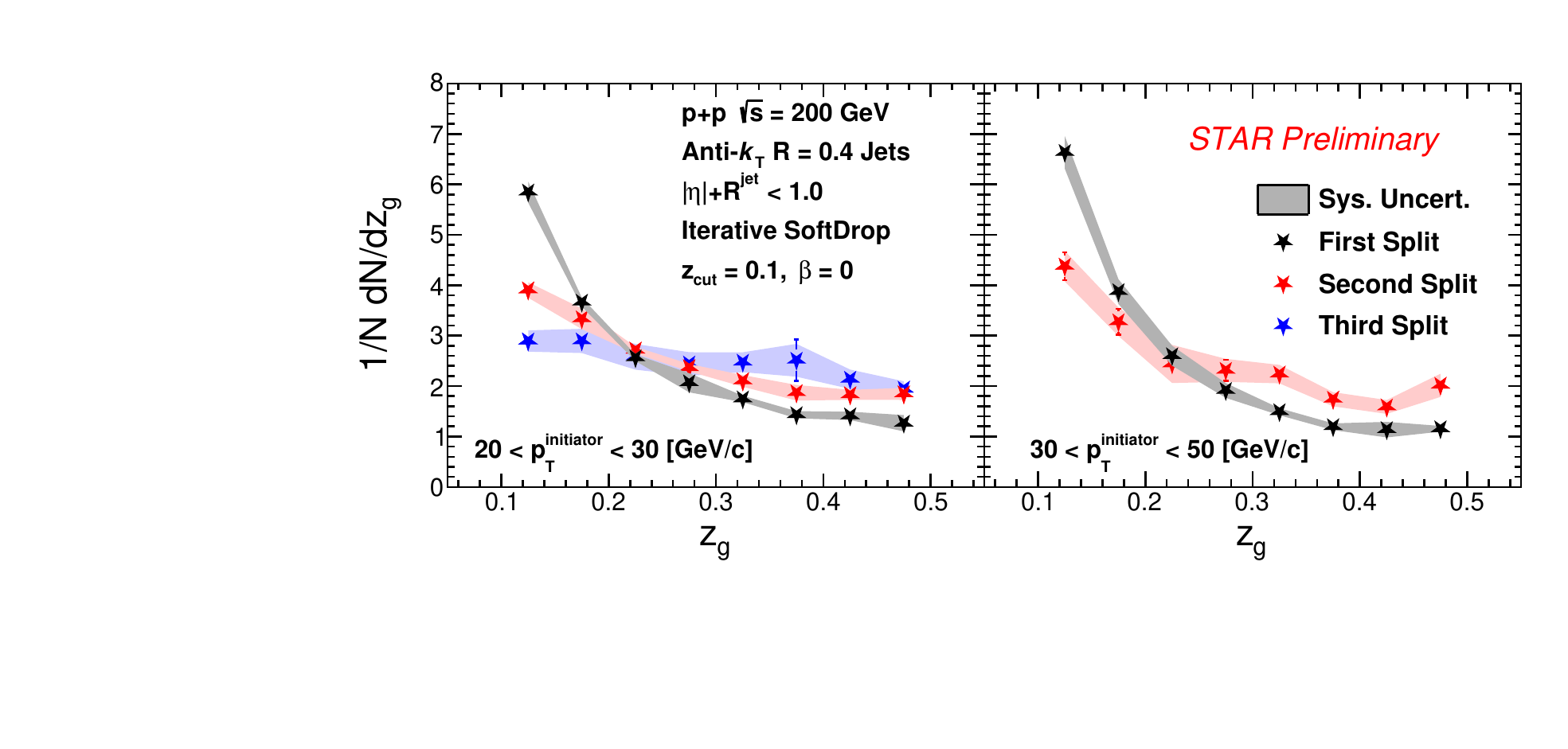}
\includegraphics[width=0.7\textwidth]{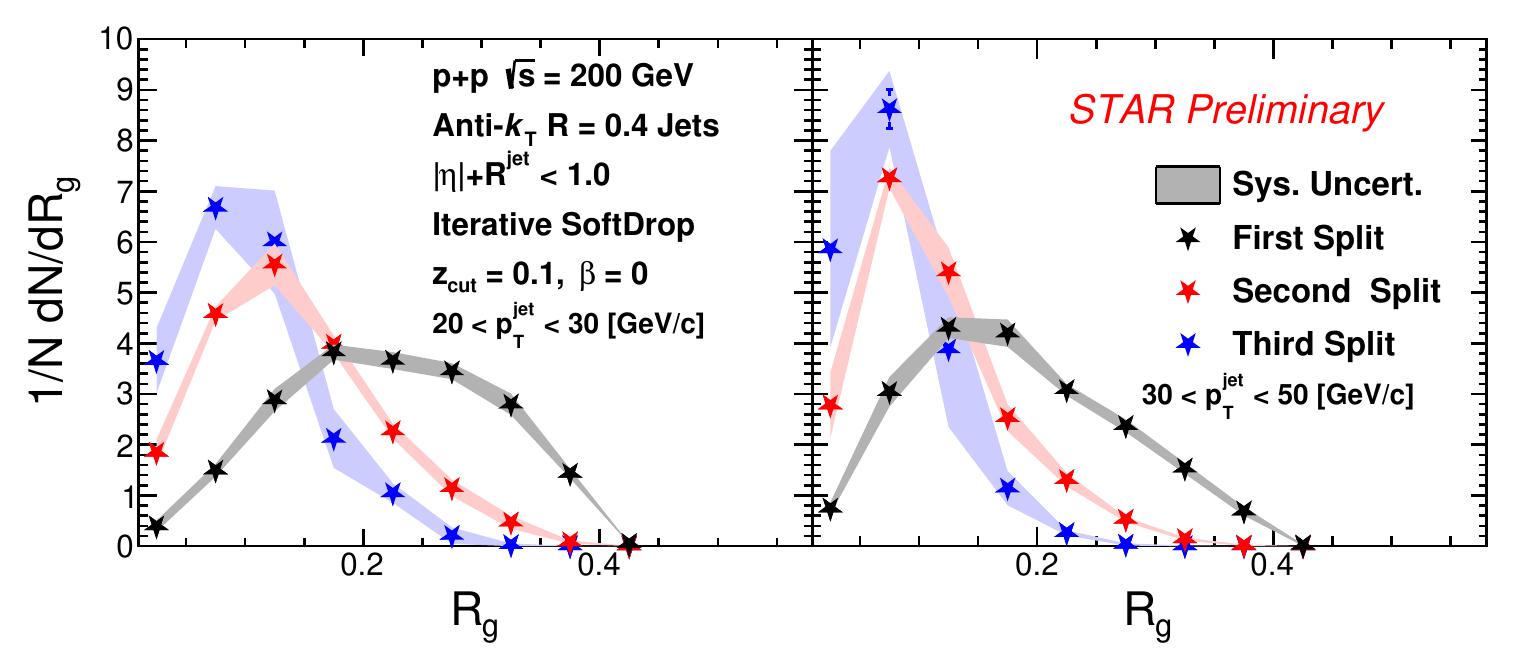}
\caption{Measurements of the iterative SoftDrop splitting observables, \zg\ (top panels) and \rg\ (bottom panels), for the first (black markers), second (red markers) and third (blue markers) splits. The top (bottom) panels are differential in initiator (jet) \pT\ for two selections corresponding to $20 < p_{\rm{T}} < 30$ (left) and $30 < p_{\rm{T}} < 50$ (right) GeV/$c$.}
\label{fig:itrsplit}
\end{figure}

The fully corrected data are shown in Fig.~\ref{fig:itrsplit} for the first, second and third splits as black, red and blue colored markers, respectively, and the shaded regions around data markers represent the total systematic uncertainty. The top panels show \zg\ for two different initiator \pT\ selections, $[20, 30]$ GeV/$c$ on the left and $[30, 50]$ GeV/$c$ on the right, and the bottom panels show \rg\ for two jet \pT\ selections. These measurements exhibit a remarkable feature of substructure evolution along the jet shower, e.g. a gradual variation in both \zg\ and \rg\ as we move from the first to the third splits. The \rg\ at a split can be interpreted as the available phase space for subsequent emissions/splits, and is also related to the virtuality at the split. As \rg\ gets progressively narrower with increasing split $n$, the shape of the \zg\ also changes from being sharply peaked at smaller values, i.e asymmetric splitting, to a flatter distribution with increased probability for symmetric splits. 

In comparing the left and right panels of Fig.~\ref{fig:itrsplit}, a weak dependence on the jet/initiator \pT\ is observed, while the phase space restrictions via selecting a split (first, second or third) significantly impacts the substructure observables. 
 
\begin{figure}[h]
\centering
\includegraphics[trim=120 60 0 20,clip,width=0.7\textwidth]{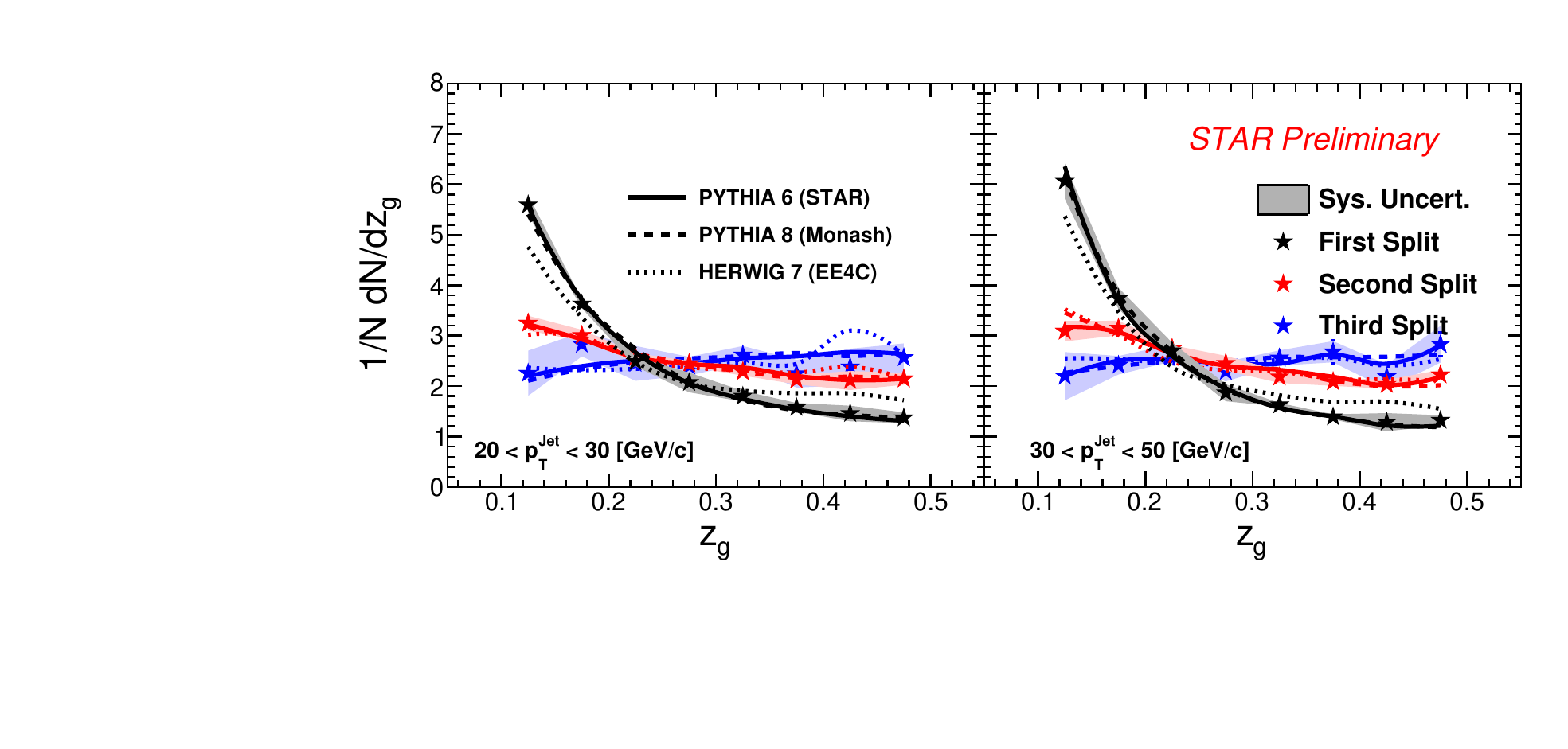}
\caption{Iterative SoftDrop \zg\ for first, second and third splits for various $p_{\rm{T, jet}}$ selections (left and right) compared to predictions from \textsc{PYTHIA 6} (solid line), \textsc{PYTHIA 8} (dashed) and \textsc{HERWIG 7} (dotted) event generators.}
\label{fig:mccomp}
\end{figure}

Figure~\ref{fig:mccomp} shows the fully unfolded \zg\ for the first (black), second (red) and third (blue) splits for $20 < p_{\rm{T, jet}} < 30$ (left) and $30 < p_{\rm{T, jet}} < 50$ (right) GeV/$c$ compared with leading order monte carlo (MC) event generators \textsc{PYTHIA 6} (solid), \textsc{PYTHIA 8} (dashed) and \textsc{HERWIG 7} (dotted). The MC models are able to reproduce the evolution of \zg\ as we increase the split $n$. The slight differences observed for the HERWIG predictions at the first split vanish for higher splits, where one expects a greater impact of non-perturbative corrections.   

\section{Conclusion}
\label{sec:conc}

STAR has measured the fully corrected iterative SoftDrop \zg\ and \rg\ distributions for the first, second and third splits along the jet clustering tree. These measurements are presented as a function of both the jet \pT\ and the initiator \pT. We observe a significant modification of the shape of \zg\ and \rg\ as we travel along the jet shower from the first to the third splits due to a constriction of the available phase space for radiations. Such an evolution can be connected to the jet's virtuality and its subsequent evolution from hard scattering scale ($Q^{2}$) to the hadronization scale ($\Lambda_{\rm{QCD}}$). The fully corrected data are compared to leading order MC event generators which showcase an overall qualitative agreement with the data albeit slight differences at the first split which are reduced for second and third splits. In the near future, the data will be compared to MC generators with varying perturbative (parton showers) and non-perturbative (hadronization, multi-parton interactions) implementations to highlight the transition between the two regions of the jet shower. This technique opens up the exciting possibility of space-time tomography in \AuAu\ collisions and enables differential measurements of jet energy loss for specific substructure.


\nolinenumbers

\end{document}